\documentclass[structabstract]{aa}  

\usepackage{graphicx}
\usepackage{natbib}
\bibpunct{(}{)}{;}{a}{}{,} 
\usepackage{txfonts}
\begin{document}
   \title{Radio observations of \object{ZwCl~2341.1+0000}: a double radio relic cluster}
 \titlerunning{ZwCl~2341.1+0000: a double radio relic cluster}

   \author{R.~J. van Weeren \inst{1}
          \and H.~J.~A. R\"ottgering\inst{1}
          \and  J. Bagchi  \inst{2}
          \and S. Raychaudhury \inst{3}
          \and H.~T. Intema \inst{1}
          \and F. Miniati \inst{4}
          \and T.~A. En\ss lin\inst{5}
          \and M. Markevitch  \inst{6}
          \and T. Erben \inst{7}
          }

   \institute{Leiden Observatory, Leiden University,
              P.O. Box 9513, NL-2300 RA Leiden, The Netherlands\\
              \email{rvweeren@strw.leidenuniv.nl}
         \and Inter-University Centre for Astronomy and Astrophysics (IUCAA), Pune 411007, India
         \and School of Physics and Astronomy, University of Birmingham, Edgbaston, Birmingham B15 2TT, England
         \and Physics Department, Wolfgang-Pauli-Strasse 27, ETH-Z\"urich, CH-8093 Z\"urich
         \and Max-Planck-Institut f\"ur Astrophysik, Karl-Schwarzschild-Str.1, PO Box 1317, 85741 Garching, Germany
         \and Harvard-Smithsonian Center for Astrophysics, 60 Garden Street, Cambridge, MA 02138
         \and Argelander-Institut f\"ur Astronomie, Universit\"at Bonn, Auf dem H\"ugel 71, D-53121 Bonn, Germany  \\
             }


 
\abstract
   {Hierarchal models of large-scale structure (LSS) formation predict that galaxy clusters grow via gravitational infall and mergers of smaller subclusters and galaxy groups. Diffuse radio emission, in the form of radio \emph{halos} and \emph{relics}, is found in clusters undergoing a merger, indicating that shocks or turbulence associated with the merger are capable of accelerating electrons to highly relativistic energies. Double relics are a rare class of radio sources found in the periphery of clusters, with the two components located symmetrically on the opposite sides of the cluster center. 
   These relics are important probes of the cluster periphery as (i) they provide an estimate of the magnetic field strength, and (ii) together with detailed modeling can be used to derive information about the merger geometry, mass, and timescale.  
    Observations of these double relics can thus be used to test the framework of LSS formation. Here we report on radio observations of  \object{ZwCl~2341.1+0000}, a complex merging structure of galaxies located at $z=0.27$, using Giant Metrewave Radio Telescope (GMRT) observations.
    }
   {The main aim of the observations is to study the nature of the diffuse radio emission in the galaxy cluster \object{ZwCl~2341.1+0000}.    }
   {We carried out GMRT $610$, $241$, and $157$~MHz continuum observations of \object{ZwCl~2341.1+0000}. The radio observations are combined with X-ray and optical data of the cluster.
   }
   { The GMRT observations show a double peripheral radio relic in the cluster \object{ZwCl~2341.1+0000}. The spectral index is $-0.49 \pm 0.18$ for the northern relic and $-0.76 \pm 0.17$ for the southern relic. We have derived values of $0.48-0.93$~$\mu$Gauss for the equipartition magnetic field strength. The relics are probably associated with outward traveling merger shock waves.
   }
   {}

   
   \maketitle

\section{Introduction}
Galaxy clusters grow by mergers of smaller subclusters and galaxy groups as predicted by hierarchical models of large-scale structure formation. During a clusters merger a significant amount of energy is released, of $10^{63}- 10^{64}$~ergs for the most massive mergers according to these models. All massive clusters have undergone several mergers in their history and presently clusters are still in the process of accreting matter. Key properties for testing models of large-scale structure (LSS) formation include the total energy budget and the detailed temperature distribution within a cluster, which are both strongly affected by the cluster's merger history. Moreover, the physics of shock waves in the tenuous intra-cluster medium (ICM) and the effect of cosmic rays on galaxy clusters are all fundamental for our understanding of LSS formation.

Diffuse radio sources with relatively steep spectra\footnote{$F_{\nu} \propto \nu^{\alpha}$, with $\alpha$ the spectral index} ($\alpha \lesssim -0.5$), which are not directly associated with individual galaxies, are observed in a number of clusters \citep[e.g., see the review by][]{2008SSRv..134...93F}. The vast majority of clusters with diffuse extended radio sources are massive, X-ray luminous and show signs of undergoing a merger. Shocks and turbulence associated with the merger are thought to be responsible for (re)accelerating electrons to highly relativistic energies and synchrotron radiation is produced in the presence of magnetic fields \citep[e.g.,][]{1998A&A...332..395E, 2000ApJ...542..608M, 2001MNRAS.320..365B, 2001ApJ...557..560P, 2003ApJ...584..190F}. The correlation between cluster mergers and diffuse radio emission is strongly suggested by current observational and theoretical work, however the connection between the diffuse radio emission and clusters mergers has still not been fully understood.

The diffuse radio sources in clusters are commonly divided into three groups: radio halos, radio mini-halos, and relics. Radio halos have smooth morphologies, are extended with sizes $\gtrsim1$~Mpc, unpolarized \citep[with upper limits of a few percent, except Abell 2255;][]{2005A&A...430L...5G}, and are found in the center of clusters. Radio halos follow the thermal X-ray emitting gas from the ICM. Radio mini-halos are not associated with merging clusters and are found in the centers of cool core clusters \citep[e.g.,][]{1991A&ARv...2..191F,  2006PhR...427....1P}. They are associated with the central, brightest cluster galaxy and have typical sizes $\lesssim 500$~kpc \citep[e.g.,][]{2009A&A...499..371G}.
Radio relics can be highly polarized, are usually found in the periphery of clusters, and have a filamentary or elongated irregular morphology. Their polarization fraction, morphology, and location can vary significantly, possibly reflecting different physical origins or conditions in the ICM; see \cite{2004rcfg.proc..335K}.

Giant radio relics are observed in the cluster periphery, with sizes up to several Mpc \citep[e.g.,][]{1991A&A...252..528G, 1994ApJ...436..654R, 2006AJ....131.2900C}. These giant radio relics sometimes show symmetric or ring-like structures and are highly polarized ($10 - 50 \%$ at 1.4 GHz). They are probably signatures of electrons accelerated at large-scale shocks. Smaller relics have been found closer to cluster centers. In this case they could be ``radio ghosts'', i.e., remnants of past AGN activity.  All known radio relics and halos are found in clusters that show signs of undergoing a merger. This gives strong support for the scenario in which the electrons are accelerated by merger-induced shocks or turbulence. \cite{2008A&A...486..347G} reported that the location of the peripheral relic in \object{Abell 521} coincided with an X-ray brightness edge. This edge could well be the shock front where particles are accelerated. The spectral index of $-1.48\pm 0.01$ for the A521 relic is in agreement with the low Mach numbers expected in clusters mergers (see below). This suggests that at least in some clusters, the relativistic particles responsible for the presence of a peripheral radio relic are accelerated in a shock front. However, not all merging clusters host a diffuse radio source, indicating that other processes must be identified for a full understanding of this phenomena.

The class of double radio relics is particular interesting because, based on current models of electron acceleration for this class of radio sources, it enables us to explore the connection between clusters mergers and shock waves \citep[e.g.,][]{1999ApJ...518..603R, 1998A&A...332..395E}. In this case two (sometimes ring-like) relics are found in the periphery, symmetric with respect to the cluster center as traced by the X-ray emission. These relics are thought to trace an outward traveling shock emanating from the cluster center, which was created during a cluster merger with a smaller substructure. An alternative shockwave-inducing mechanism is that of external ``accretion'' shocks where the  intergalactic medium (IGM) in filaments of galaxies funnels deep into the cluster \citep{2000ApJ...542..608M, 2003MNRAS.342.1009M, 2003ApJ...585..128K}. The merger shocks are weaker than external accretion shocks, since the gas has already been heated by these external shocks. The external accretion shocks occur farther out than the merger shocks up to a few times the virial radius of the cluster. However, the gas density is much higher closer to the cluster center so the energy densities in cosmic ray (CR) electrons and magnetic fields are also higher. Accretion shocks of filaments can therefore also be responsible for the occurrence of peripheral radio relics \citep{2001ApJ...562..233M}. While double relics are allowed in the filament accretion picture, they are not necessarily symmetric with respect to the X-ray elongation axis. A symmetric configuration  arises naturally in the binary cluster merger picture \citep[e.g.,][]{1999ApJ...518..603R}.

Double relics are rare, with only five of them known in total. \object{Abell 3667} hosts a giant double relic \citep{1997MNRAS.290..577R} with a total size of 3.8~Mpc. A numerical model for the diffuse radio and X-ray emission in A3667 was presented by \cite{1999ApJ...518..603R} where the formation of the double relics was explained by an outgoing merger shock front. The model suggested that the two relics were produced by a slightly off-axis merger about 1~Gyr ago, with the merging subcluster having 20\% of the mass of the primary cluster. Their model correctly predicted (i) the orientation of the X-ray emission, (ii) the spectral index gradients of the radio relics with steeper spectra farther away from the shock front due to spectral aging, (iii) the approximate location of the radio emission with respect to the X-ray emission, and (iv) roughly the shape of the two relics. 

The double radio relic around \object{Abell 3376}  \citep{2006Sci...314..791B} has a maximum size of about 2~Mpc. \cite{2006Sci...314..791B} argue that the relics can be explained by outgoing merger shocks or by accretion shocks tracing the infall of the IGM or warm-hot intergalactic medium (WHIM) onto the cluster at approximately the viral radius. \object{RXC J1314.4$-$2515} \citep[][]{2002A&A...392..795V, 2005A&A...444..157F, 2007A&A...463..937V} also hosts a double relic. This is the only cluster known to have a radio halo and a double radio relic.

\cite{2009A&A...494..429B} present observations of double relics in \object{Abell 2345} and \object{Abell 1240}. The polarization fraction for the relics in \object{A2345} was found to be 22\% and 14\%. For \object{A1240}, values of 26\% and 29\% were derived. Steepening of the spectral index was observed away from the shock front towards the center of the cluster for one relic in \object{A1240} (the data for the other relic was consistent with this trend), as well as for one relic in \object{A2345}. They concluded that these trends are consistent with shock model predictions. An opposite trend has been seen for the second relic in \object{A2345}, i.e., steeper spectra farther away from the cluster center. This trend could be explained by the peculiar position of the relic between two merging substructures.

In this paper we present deep low-frequency Giant Metrewave Radio Telescope (GMRT) observation of the merging system \object{ZwCl~2341.1+0000} at 610, 241, and 157~MHz. These observations were taken to investigate the nature of the diffuse radio emission within the cluster and explore the connection between the radio and the X-ray emission. \object{ZwCl~2341.1+0000} ($\alpha=23^{h}43^{m}40^{s}$, $\delta=00^{\circ}16\arcmin39\arcsec$) is a complex cluster galaxies, composed out of several different subclusters, probably at the junction of supercluster filaments \citep[\citeauthor{temple}~in~prep.; ][]{2002NewA....7..249B}. The system is located at a redshift of $z = 0.27$ (using SDSS DR7 spectroscopic redshifts of several galaxies in the vicinity). The cluster is also listed as \object{SDSS~CE~J355.930756+00.303606} and \object{NSCS~J234339+001747}. Galaxy isodensity maps, derived from SDSS imaging data, show an elongated cluster of galaxies, including several subclusters distributed roughly along a north-south axis. A galaxy filament is seen branching off from the main structure towards the northeast.

\cite{2002NewA....7..249B} discovered the presence of diffuse radio emission in 1.4~GHz NVSS images \citep[FWHM 45\arcsec][]{1998AJ....115.1693C}, as the emission was not seen in higher resolution 5\arcsec~FIRST images \citep{1997ApJ...475..479W, 2003yCat.8071....0B} around \object{ZwCl~2341.1+0000}.
Very Large Array (VLA) observations at 325~MHz confirmed the presence of diffuse emission in the cluster, although the resolution in the corresponding images was rather low ($108\arcsec$), making it difficult to disentangle the diffuse emission from compact sources. At 325~MHz an extension of diffuse emission towards the northeast was suggested, with a spectral index $< -0.9$, as well diffuse radio emission following the north-south galaxy distribution. The spectral index (between 1400 and 325~MHz) of the diffuse emission was found to be $\sim$$-0.5$, which is not steep compared to other diffuse radio sources in clusters. The derived equipartition magnetic field strength was $0.3-0.5\mu$Gauss. \citeauthor{2002NewA....7..249B} also found the system to be an extended source in ROSAT PSPC All-Sky X-ray Survey. They concluded that the diffuse radio emission was the first evidence of cosmic-ray particle acceleration taking place at cosmic shocks in a magnetized IGM on scales $\ge 5$~Mpc.

\begin{figure*}
    \begin{center}
      \includegraphics[angle = 90, trim =0cm 0cm 0cm 0cm,width=1.0\textwidth]{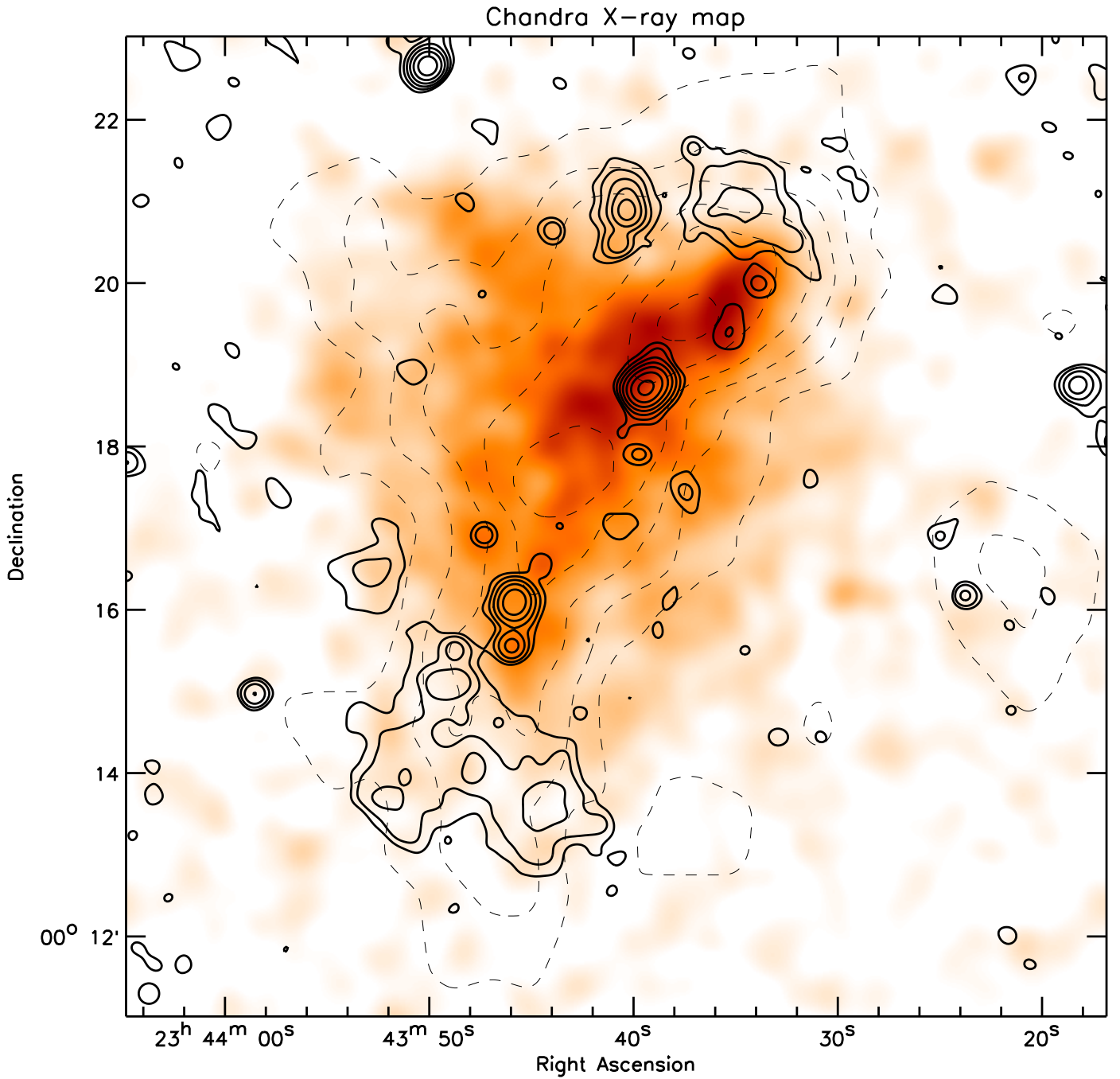}
       \end{center}
      \caption{X-ray emission from Chandra in the $0.5- 3.0$~keV energy band. Point sources were excluded from the image, see \citeauthor{temple} (in prep.). The image has been convolved with a circular Gaussian of 24\arcsec. The solid contours represent the radio emission at 610~MHz from the GMRT. The radio map has been convolved to a circular beam of 15\arcsec~to better show the diffuse radio emission. The radio contours are drawn at $[1, 2, 4, 8, 16, 32,  ...]  \times 0.224$ mJy beam$^{-1}$. Dashed contours show the galaxy isodensity contours from SDSS, corresponding roughly to a limit of $M_{\mathrm{V}} = -18.1$ (i.e., $M^\ast + 2.4$). The dashed contours are drawn at $[2, 3, 4, 5,...]$ galaxies arcmin$^{-2}$, using the redshift cuts described in the text.} 
            \label{fig:xray}
 \end{figure*}

\citeauthor{temple} (in prep.) obtained high sensitivity X-ray observations of \object{ZwCl~2341.1+0000} using the Chandra and XMM-Newton satellites. They find the X-ray emission span about 3.3~Mpc in the north-south direction. An extension towards the east is also visible (see Fig.~\ref{fig:xray}). The X-ray data show a complex, clearly disturbed ICM, indicating a merger in progress.

The layout of this paper is as follows. In Sect. \ref{sec:obs-reduction} we give an overview of the observations and data reduction. In Sects. \ref{sec:results} and \ref{sec:xray} we present the radio maps and compare them to the X-ray and galaxy distribution. In Sect. \ref{sec:alphaandgauss} we present the spectral index maps and derive the equipartition magnetic field strength. We end with a discussion and conclusions in Sects.~ref{sec:discussion} and \ref{sec:conclusion}.

Throughout this paper we assume a $\Lambda$CDM cosmology with $H_{0} = 71$ km s$^{-1}$ Mpc$^{-1}$, $\Omega_{m} = 0.27$, and $\Omega_{\Lambda} = 0.73$. At a distance of $z=0.27$, 1\arcmin~corresponds to a physical scale of 246~kpc.

\section{Observations \& data reduction}
\label{sec:obs-reduction}

High-sensitivity radio observations of the merging cluster \object{ZwCl~2341.1+0000} were carried out with the GMRT at 610, 241, and 157 MHz. 
The 241 and 610~MHz observations were carried out simultaneously using the dual-frequency mode. At 610~MHz only the righthanded circular polarization (RR) was recorded and at 241~MHz only the left-handed circular polarization (LL).  Both upper (USB) and lower (LSB) sidebands were recorded at 610~MHz resulting in a total bandwidth of 32~MHz. At 241~MHz only the USB was recorded with a effective bandwidth of 6~MHz (see below). The data were collected in spectral line mode with $128$ channels per sideband (IF), resulting in a spectral resolution of $125$~kHz per channel. The 157~MHz observations recorded the LSB sideband and included both LL and RR polarizations. The data were also collected in spectral line mode with $128$ channels and a total bandwidth of 8~MHz (62.5~kHz per channel). The total integration time for the dual-frequency 610 and 241~MHz observations was 12.9~hours. For the 157~MHz observations this was 6.25~hours. A summary of the observations is given in Table~\ref{tab:observations}.

\begin{table*}
\begin{center}
\caption{Observations}
\begin{tabular}{llll}
\hline
\hline
& 610 MHz & 241 MHz & 157 MHz \\
\hline
Bandpass \& flux calibrator(s) & 3C 48, 3C 147 & 3C 48, 3C 147 & 3C 48, 3C 286 \\
Phase calibrator(s)                   & 2225-049 &2225-049 & 0025-260 \\
Bandwidth		  & 32 MHz& 6 MHz$^{a}$ & 8 MHz \\
Channel width                      &125 kHz & 125 kHz &  62.5 kHz\\
Polarization			& RR& LL & RR+LL \\
Sidebands			& USB+LSB & USB & USB \\
Observation dates				& 3, 4, 5, 6 Sep 2003& 3, 4, 5, 6 Sep 2003 & 2, 3 Jun 2006 \\Integration time per visibility               	& 16.9 s &16.9 s& 16.9 s \\	
Total duration$^{b}$		& 12.90 hr& 12.90 hr& 6.25 hr\\
Beam size$^{c}$			&  $6.9\arcsec \times 4.3\arcsec$ & $17.5\arcsec \times 10.3\arcsec$ & $21.1\arcsec \times 17.1\arcsec$ \\
rms noise ($\sigma$)	& 28 $\mu$Jy beam$^{-1}$ & 297 $\mu$Jy beam$^{-1}$ & 1.36 mJy beam$^{-1}$\\	
\hline
\hline
\end{tabular}
\label{tab:observations}
\end{center}
$^{a}$ the total bandwidth is 16~MHz but only 6~MHz is usable due to RFI\\
$^{b}$ on-source time (not including calibrators and slewing time)\\
$^{c}$ restoring beam, robust weighting parameter was set to $0.5$ \citep{briggs_phd} 
\end{table*}

\begin{figure*}
    \begin{center}
      \includegraphics[angle = 90, trim =0cm 0cm 0cm 0cm,width=1.0\textwidth]{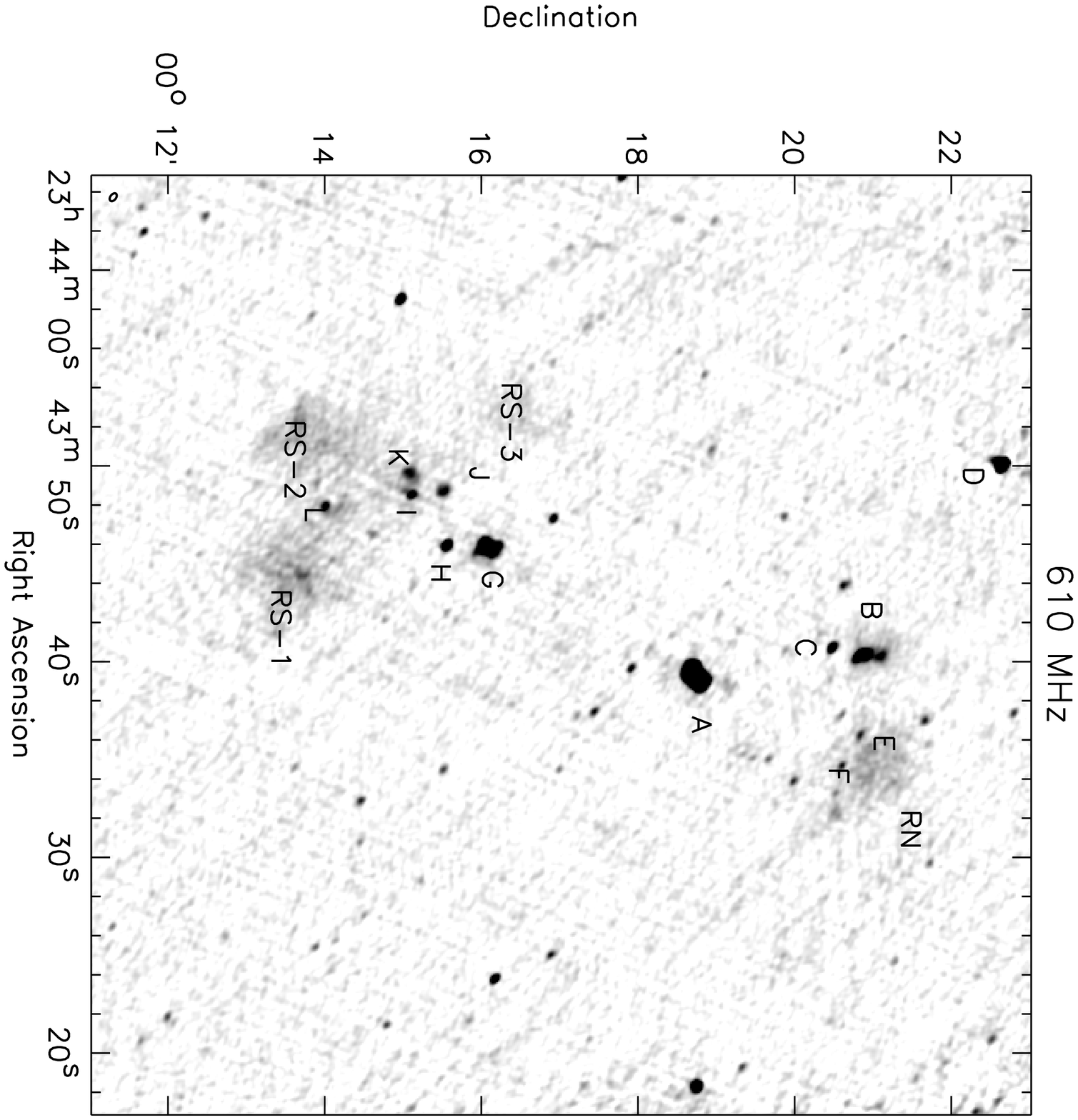}
       \end{center}
      \caption{Radio map at 610~MHz. The greyscale image represent the high-resolution map with a restoring beam size of $6.9\arcsec \times 4.3\arcsec$, indicated in the bottom left corner. }
                  \label{fig:map610}
       \end{figure*}
The data was reduced and analyzed with the NRAO Astronomical Image Processing System (AIPS) package.  Bandpass calibration was carried out using the standard flux calibrators: 3C48, 3C147, and 3C286. The flux densities for our primary calibrators were taken from the \cite{perleyandtaylor} extension to the \cite{1977A&A....61...99B} scale. This results in $29.43$ Jy for 3C48 and $38.26$ Jy for 3C147 at $610$~MHz; $51.05$ Jy and $59.42$ Jy at $241$~MHz. At $157$~MHz, the flux density scale gives $62.71$~Jy for 3C 48 and 31.02~Jy for 3C 286. A set of channels (typically 5) free of radio frequency interference (RFI) were taken to normalize the bandpass for each antenna. We removed strong radio frequency interference (RFI) automatically with the AIPS task `FLGIT', but also carefully visually inspected the data for remaining RFI using the AIPS tasks `SPFLG' and `TVFLG'. The RFI was partially strong at 241 and 157~MHz. The antenna gains were set using the primary (bandpass) and secondary calibrators and applied to our target source. We have not chosen to average any channels to aid further removal of RFI at a later stage. The first and last few channels of the data were discarded as noisy. At 241~MHz, only the first $\sim$$50\%$ of the 16~MHz bandwidth is usable. 

For making the images (used subsequently in the selfcalibration), we used the polyhedron method \citep{1989ASPC....6..259P, 1992A&A...261..353C} to minimize the effects of non-coplanar baselines. At 610~MHz the USB and LSB were reduced separately. We used a total of 199 facets covering $\sim$$2$ times the full primary beam in order to remove sidelobes from strong sources far away from the field center. After a first round of imaging, ``ripples'' in the background were subsequently removed by identifying corresponding bad baseline(s). Several rounds of phase selfcalibration were carried out before doing a final amplitude and phase selfcalibration. A deep image was made, and the corresponding clean component model was subtracted from the UV-data. This data was then visually inspected for remaining low-level RFI and flagged automatically with `FLGIT' using a $6\sigma$ rms clip before adding back the clean component model to the UV-data. The data was then averaged to 22 (610~MHz),  8 (241~MHz), and 11 (157~MHz), channels. The 610~MHz USB and LSB  were combined using the tasks `UVFLP'\footnote{http://www.mrao.cam.ac.uk/\~{}dag/UVFLP/} \citep{2007MNRAS.376.1251G} written by D. A. Green, `BLOAT', and `DBCON' so they could be simultaneously imaged and cleaned.

The 157~MHz data was further calibrated for ionospheric phase distortions, because they can become quite severe at this frequency, using the SPAM package by \cite{2009A&A...501.1185I}. Phase solutions of the 10 brightest sources within the field of view were used to fit an ionospheric model to the data. The resulting direction-dependent phase corrections were applied during imaging using the Cotton-Schwab clean algorithm \citep{1984AJ.....89.1076S, 1999ASPC..180..357C, 1999ASPC..180..151C}. 
 This lowered the rms noise in the image by factor of $1.2$ with respect to conventional selfcalibration. The final 610 and 241~MHz images were made using CASA (formerly AIPS++)\footnote{http://casa.nrao.edu/}, using w-projection \citep{2005ASPC..347...86C,2008ISTSP...2..647C} with $512$ internal planes. \footnote{The 157~MHz images could not be made with the CASA imager since the direction-dependent phase corrections could only be applied using specialized imaging routines from the SPAM package.} We weighted our UV-data using robust weighting $0.5$ \citep{briggs_phd}, increasing the noise level in the maps by about 15\%.  
The $610$ and $241$~MHz were cleaned with the Multi-scale CLEAN algorithm \citep{2008ISTSP...2..793C} yielding significantly better results than Clark CLEAN for large-scale diffuse emission. Five different convolving scales were used at $610$~MHz (0.0\arcsec, 3.6\arcsec, 18.0\arcsec, 60.0\arcsec, 120.0\arcsec) and three at $241$~MHz (0.0\arcsec, 9.3\arcsec, 15.5\arcsec). We checked absolute flux calibration by obtaining flux measurements from the literature for 15 strong compact sources within our field from 74~MHz to 4.8~GHz. These flux measurements were fitted with second order polynomials in log-space to obtain the radio spectra. The fitted spectra were then compared to fluxes measured from the GMRT maps. The accuracy of the absolute flux calibration was found to be about $5-10\%$ at all three frequencies, in agreement with values derived by \cite{2004ApJ...612..974C}.

The theoretical thermal noise in the image is given by\footnote{http://www.gmrt.ncra.tifr.res.in/gmrt\_hpage/

Users/doc/manual/UsersManual/node13.html}
\begin{equation}
\sigma_{\mathrm{thermal}} = \frac{\sqrt{2}T_{\mathrm{sys}}}{G\sqrt{n(n-1) N_{\mathrm{IF}} \Delta\nu t_{\mathrm{int}} }} \mbox{ ,}
\end{equation}
with the $T_{\mathrm{sys}}$ the system temperature ($T_{\mathrm{sys}} = T_{\mathrm{R}} + T_{\mathrm{sky}} + T_{\mathrm{ground}}$, with $T_{\mathrm{R}}$, $T_{\mathrm{sky}}$, and $T_{\mathrm{ground}}$ the receiver, sky, and ground temperatures respectively), $G = 0.32$ K Jy$^{-1}$ the antenna gain, $n \approx 26$ the number of working antennas, $N_{\mathrm{IF}}$  the number of sidebands used (both recording RR and LL polarizations), $\Delta\nu$ the bandwidth per sideband, and $t_{\mathrm{int}}$ the net integration time. At 610~MHz $T_{\mathrm{sys}} = 92$ K, $N_{\mathrm{IF}}=1$\footnote{There are two sidebands but only one polarization is recorded}, $\Delta\nu = 13.5$~MHz, and $t_{\mathrm{int}} = 12.9$ hrs. The expected thermal noise for $12.9$ hrs integration time is about $22$ $\mu$Jy beam$^{-1}$, where we take into account that typically $25\%$ of the data is flagged due to RFI. The noise level in our maps is 28$\mu$Jy beam$^{-1}$, which is close to the thermal noise since weighting increased our noise level by about 10\%. The system temperature at 157~MHz is $482$ K and $177$~K at 241~MHz. In both cases $T_{\mathrm{sky}}$ dominates the contribution over $T_{\mathrm{R}}$  (at all frequencies $T_{\mathrm{ground}} < T_{\mathrm{R}}$). The thermal noise at $241$~MHz ($\Delta\nu = 6.9$~MHz, $N_{\mathrm{IF}}=0.5$, $t_{\mathrm{int}} = 12.9$ hrs) is expected to be 96 $\mu$Jy beam$^{-1}$.  At 157~MHz we expect $245$\noindent $\mu$Jy beam$^{-1}$ ($\Delta\nu = 6.0$~MHz, $N_{\mathrm{IF}}=1$, and $t_{\mathrm{int}} = 6.25$ hrs). The noise levels in our 241 and 157~MHz images are 297 $\mu$Jy beam$^{-1}$ and 1.36 mJy beam$^{-1}$, respectively. These are a factor $3.1$ and $5.5$ higher than the expected thermal noise. This is probably caused by remaining RFI, pointing errors, and other phase/amplitude errors, which cannot be solved for in our calibration. A significant amount of RFI is still present on short baselines at 157~MHz. By flagging these baselines, the noise level decreases to about $0.97$ mJy beam$^{-1}$. We have, however, not chosen to use this map as the diffuse emission from the cluster is removed together with the RFI.

\section{Results}
\label{sec:results}

\begin{figure}
    \begin{center}
       \includegraphics[angle = 90, trim =0cm 0cm 0cm 0cm,width=0.5\textwidth]{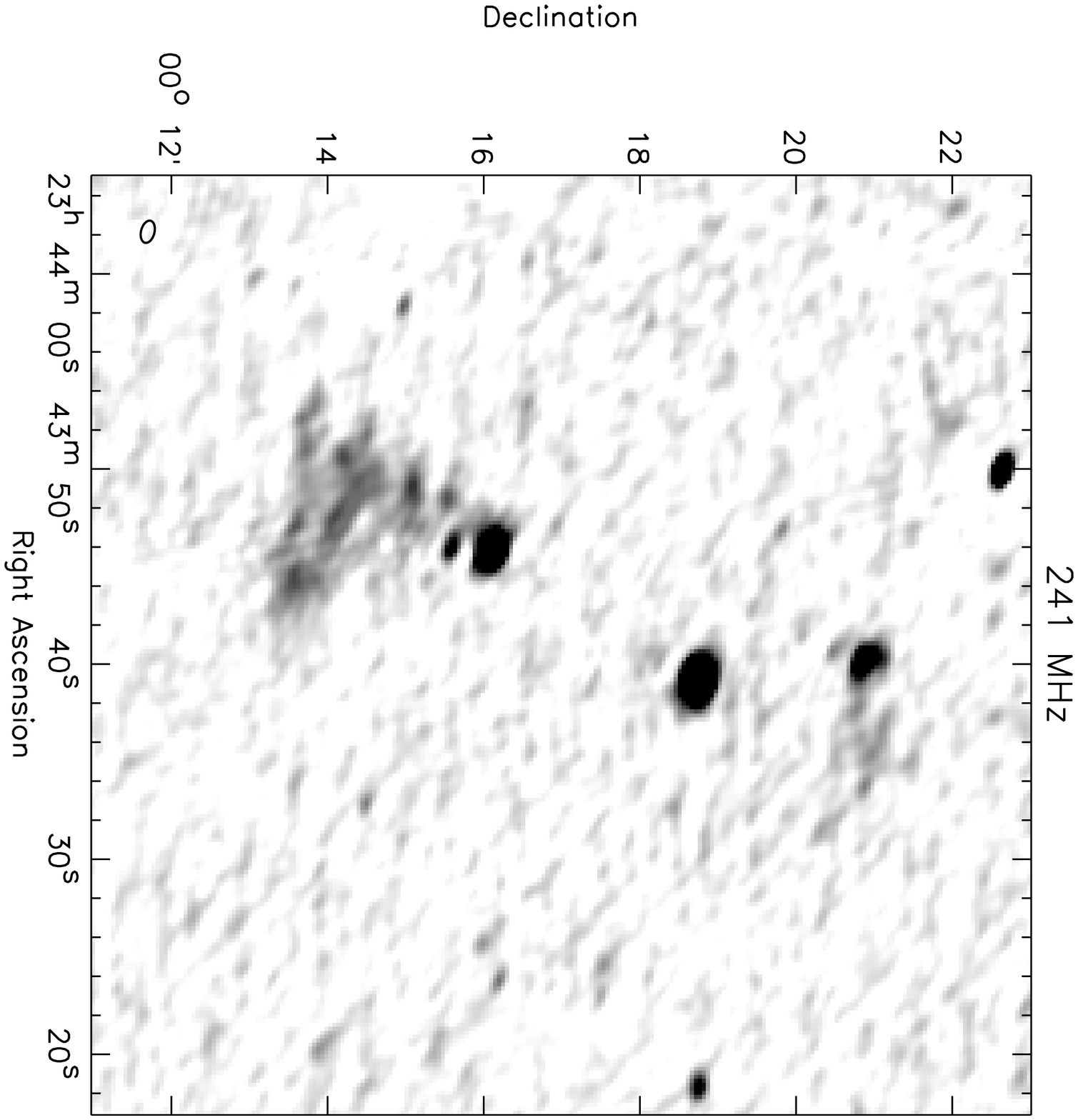}
       \end{center}
      \caption{Radio map at 241~MHz. The restoring beam size is $17.4\arcsec \times 10.3\arcsec$.}
            \label{fig:map235}
 \end{figure}

\begin{figure}
    \begin{center}
      \includegraphics[angle = 90, trim =0cm 0cm 0cm 0cm,width=0.5\textwidth]{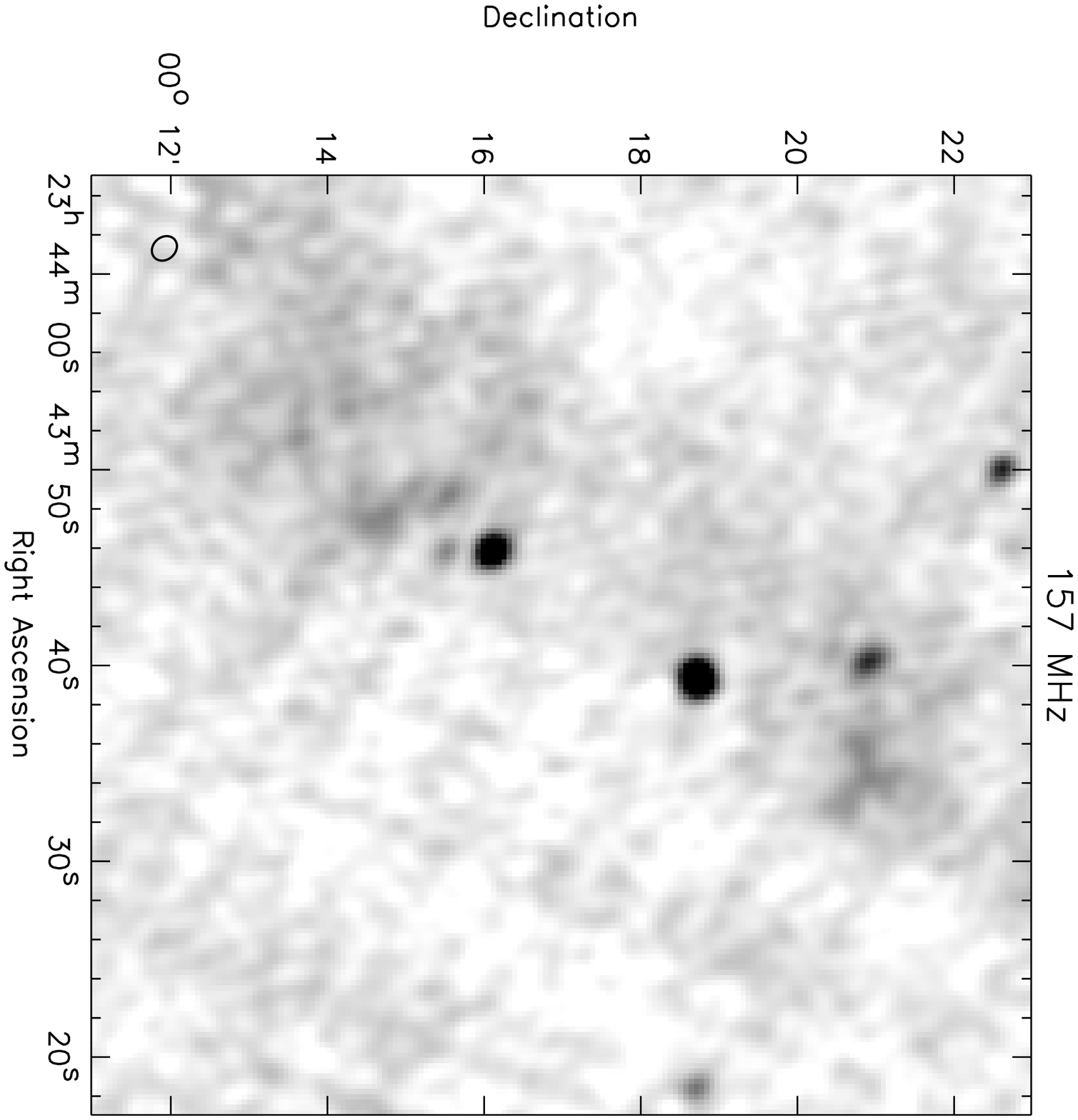}
       \end{center}
             \caption{Radio map at 157~MHz.  The restoring beam size is $21.1\arcsec \times 17.1\arcsec$.}
             \label{fig:map150}
 \end{figure}

A summary of the maps made, the corresponding restoring beam values and noise levels, is given in Table~\ref{tab:observations}. The high-resolution maps at three different frequencies are shown Figs. \ref{fig:map610}, \ref{fig:map235}, and \ref{fig:map150}.

We labeled the compact ($\lesssim 30\arcsec$) radio sources alphabetically in order to clarify the discussion of the different sources found in the maps, see Fig.~\ref{fig:map610} and Appendix \ref{sec:compact}. We only chose to label the brightest sources and those located close to regions with diffuse emission. The 610~MHz, as well as the 241~MHz image, show two regions of diffuse emission, one to the north of the cluster center (RN: relic north) and one in the south (RS: relic south). The southern diffuse region consists of three different components labeled RS-1, RS-2, and RS-3. RS-3 seems to be detached from RS-1 and RS-2. A hint of the diffuse emission is visible in the 157~MHz map at the $3\sigma$ level. Given that the diffuse sources are located in the periphery of the cluster and the diffuse emission does not seem to be associated with any of the radio galaxies, we will refer to them as relics (see Sect. \ref{sec:discussion}).  Several compact sources related to AGNs, including several possible head-tail sources, are located within the cluster. Other compact sources are related to AGNs located behind or in front of the cluster.  An overview of the labeled sources with their flux densities and spectral indices is given in Table~\ref{tab:sources}.

\begin{table*}
\begin{center}
\caption{Source properties}
\begin{tabular}{lllllllll}
\hline
\hline
Source & RA$^{a}$ & DEC & $F_{610}$ & $F_{241}$ & $F_{157}$ & $\alpha^{b}$ & resolved/unresolved & morphology/type \\                   
                          &   &  & mJy & mJy & mJy \\                                                                                               
\hline                                                                                                                                                               
A & 23 43 39.3 & 00 18 43.6 & $29.6 \pm 2.9$   & $69.4\pm 7.0$  & $69.7 \pm 7.8$&$-0.92 \pm -0.15$& resolved & head-tail\\   
B & 23 43 40.3 & 00 20 52.8 & $10.1 \pm 1.2$   & $20.4 \pm 2.6$ &  $38.5 \pm 7.6$ &$-0.76 \pm 0.19$ &resolved & head-tail\\    
C & 23 43 40.7 & 00 20 29.3 & $1.53 \pm 0.18$ & \ldots &    \ldots  &   \ldots& unresolved & \ldots \\                                 
D & 23 43 50.1 &  00 22 39.7 & $12.9 \pm 1.2$  &  $15.8 \pm 1.6$ & $19.5 \pm 3.0$ &$-0.22 \pm 0.15$& unresolved & \ldots\\    
E & 23 43 36.2 & 00 20 51.0 & $0.79 \pm 0.12$ & \ldots &    \ldots  &   \ldots & unresolved & \ldots\\ 
F & 23 43 34.6 &  00 20 36.5 & $0.66 \pm 0.11$ & \ldots &    \ldots  &   \ldots & unresolved & \ldots\\ 
G & 23 43 45.8 & 00 16 05.8 & $10.2 \pm1.2$     & $42.2 \pm 4.7$ & $53.2 \pm 5.3$ & $-1.53 \pm 0.17$& resolved & head-tail\\   
H & 23 43 46.0 & 00 15 33.5 & $2.33 \pm 0.25$  &  $6.36 \pm 1.2$ &    \ldots  & $-1.08 \pm 0.23$& unresolved & \ldots\\ 
I & 23 43 48.6   &   00 15 06.5 & $1.30 \pm 0.37$ & \ldots &    \ldots  &   \ldots & unresolved & \ldots   \\ 
J & 23 43 48.7  &  00 15 30.9 & $1.44 \pm 0.25$ & \ldots &    \ldots  &   \ldots &resolved & \ldots \\ 
K & 23 43 49.6  & 00 15 05.5 & $1.76 \pm 0.43$  & \ldots &    \ldots  &   \ldots &resolved & \ldots\\ 
L & 23 43 47.9  &  00 14 00.4 & $1.33 \pm 0.32$ & \ldots & \ldots & \ldots& unresolved & \ldots\\ 
RN & $\sim$23 43 35& $\sim$00 21 00& $14 \pm 3$ & $21 \pm 7$ & $\sim$$36$&  $-0.49 \pm 0.18^{c} $& resolved & diffuse, relic\\
RS-1 + RS-2 &  $\sim$23 43 50  & $\sim$00 14 20&$37 \pm 13$& $72 \pm 21$ & $\sim$$70$  &$-0.76 \pm 0.17^{c} $ & resolved &diffuse, relic\\
RS3           &  $\sim$23 43 52  & $\sim$00 16 15&$6 \pm 3$& \ldots & \ldots  & \ldots & resolved & diffuse, relic \\
\hline
\hline
\end{tabular}
\label{tab:sources}
\end{center}
$^{a}$ position (peak flux) derived from the 610~MHz map\\
$^{b}$ spectral index between 610 and 241~MHz\\
$^{c}$ Spectral index was calculated using our fluxes at 610, 241, and 157~MHz, combined with the fluxes at 325 and 1400~MHz from \cite{2002NewA....7..249B}, see Sect. \ref{sec:spectra}\\
\end{table*}

Source A is located in the center of the clusters (near the X-ray emission peak, see Sect. \ref{sec:xray}). Other bright sources include B northeast to RN, D in the far north, and G  in the south north of RS-2. Several fainter sources are visible close to relic RN, namely E and F, and nearby RS, namely H, I, J, K, and L. 

Sources A, B, and D are also detected by the 1.4 GHz FIRST survey. Source G is not listed in the FIRST catalog, but a hint is seen in the FIRST image.  In the 1.4 GHz NVSS image sources A, B, D, and G are visible, and the combined emission from sources I, J, K, L, RS-2, and RS-1.  A hint of source RN is seen (blended with B). In the 325~MHz image from \cite{2002NewA....7..249B} source A and B are visible, although the resolution is not high enough to separate source B from the relic RN. The northeast extension of diffuse emission towards source D, which was detected at the $2\sigma$-level in the 325~MHz image from \citeauthor{2002NewA....7..249B}, is not visible in any of our maps, indicating that this feature is not real.

The deep high-resolution 610~MHz map shows that the radio relics RN and RS are truly diffuse and do not originate from compact sources. RN has a size of about 1.5\arcmin~by 1\arcmin~($0.37$ by $0.25$~Mpc) and a total flux of  14, 21, and 36 mJy at 610,  241, and 157~MHz, respectively. RS (RS-1 + RS-2 + RS-3) has a size of about 5\arcmin~by 2\arcmin~($1.2$ by $0.49$~Mpc) and a total flux of 43 mJy (610~MHz), 72 (241~MHz), and 70 mJy (157~MHz), see Table~\ref{tab:sources}.  We subtracted the flux contribution of the compact sources (E, F, I, J, K, and L) at 610~MHz from RS and RN. At 241 and 157~MHz we assumed a spectral index of $-0.5$, because the resolution at those frequencies was too low to properly identify these sources within the diffuse emission. The flux of RN and RS was determined by measuring the flux within a region showing the diffuse emission (using the lowest contour visible in the convolved 610~MHz image in Fig.~\ref{fig:xray}) with the AIPS task `TVSTAT'. 

A description of the compact sources and their optical counterparts is given in Appendix \ref{sec:compact}.


\section{Radio, X-ray, and galaxy distribution comparison}
\begin{figure}
    \begin{center}
      \includegraphics[angle = 90, trim =0cm 0cm 0cm 0cm,width=0.5\textwidth]{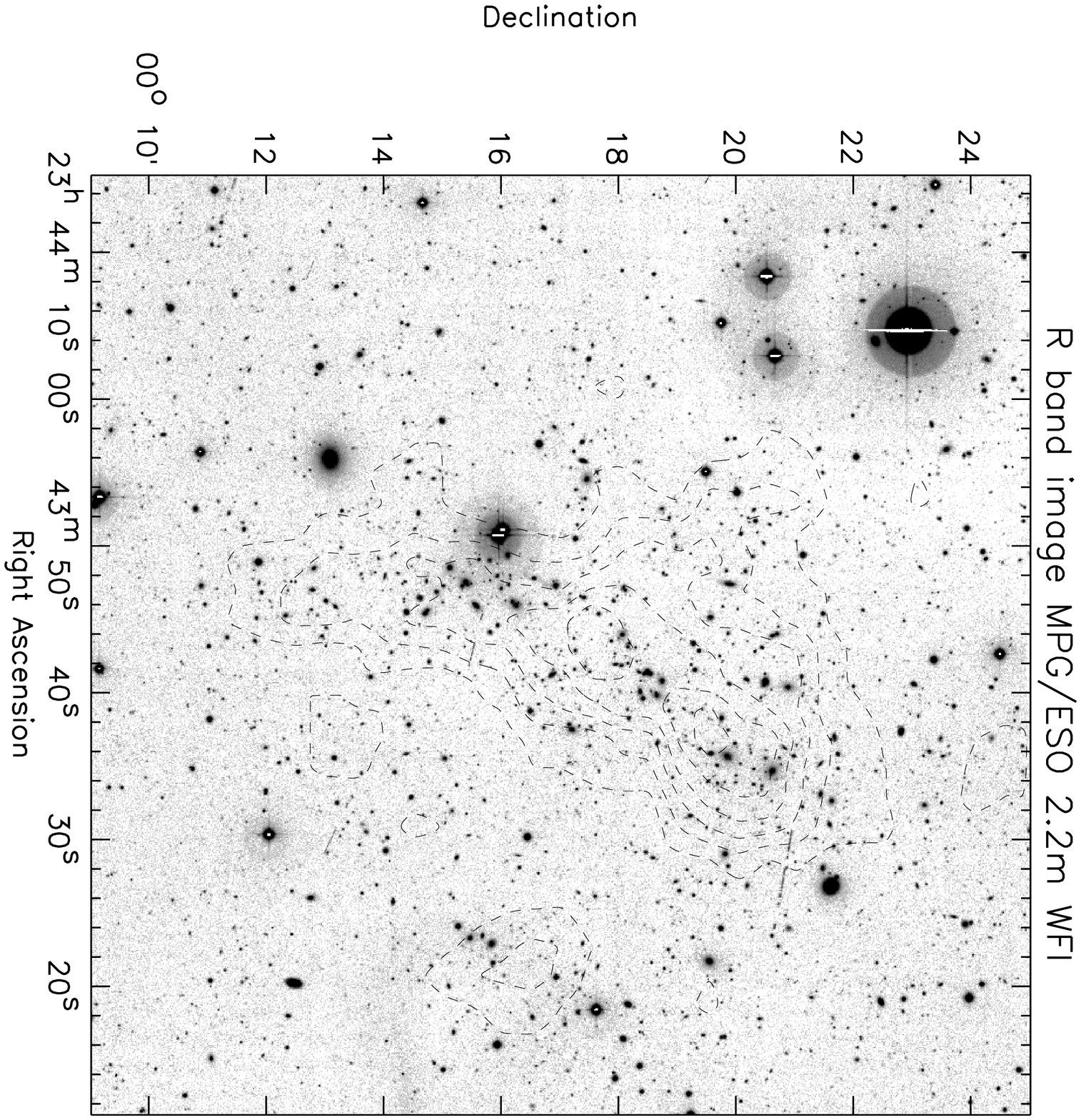}
       \end{center}
             \caption{Large-scale galaxy distribution around \object{ZwCl~2341.1+0000}. The optical R band image, a combination of several individual exposures resulting in a total exposure time of  7800~s, was obtained with the WFI at the MPG/ESO 2.2m telescope. Dashed contours show the galaxy isodensity contours from SDSS, for details see Fig.~\ref{fig:xray}. The image shows an area of $16\arcmin \times 16\arcmin$ ($\sim$ 4~Mpc~$\times$~4~Mpc) in size.}
             \label{fig:erben}
 \end{figure}
\label{sec:xray}
An X-ray map of $0.5- 3.0$~keV X-ray emission from Chandra with radio contours is shown in Fig.~\ref{fig:xray}. The  Chandra image was taken from \citeauthor{temple} (in prep). These observations had a net exposure time of $29.5$ ks, and the cluster was placed on the ACIS I3 CCD array. Point sources were removed from the maps based on the criteria described in \citeauthor{temple}. The X-ray image shows extended emission roughly in the north-south direction. A fainter northeastern extension is also visible. The morphology clearly shows a highly distorted ICM, an indication of a recent or ongoing merger. The global temperature in the cluster was found to be $\sim5$~keV. The cluster has a total X-ray luminosity ($L_{\mathrm{X}}$, $0.3-8.0$~keV) of $3 \times 10^{44}$~erg~s$^{-1}$. The combined (restframe) radio power $P_{1.4~\mathrm{GHz}}$ of the two relics is $\sim 5\times 10^{24}$ W~Hz$^{-1}$. The two diffuse radio structures are located to the north and the south of the X-ray emission, symmetrically with respect to the cluster center. The radio emission is located outside the X-ray bright area, similar to the double relics found in A3667, A3376, A2345, and A2140 \citep{1997MNRAS.290..577R, 2006Sci...314..791B, 2009A&A...494..429B}. The relics appear elongated perpendicular to the direction of the main merger apparent from the X-ray data (the main merger axis is orientated north-south). No central radio halo is detected, with a limit on the surface brightness of $\sim1.5$~$\mu$Jy~arcsec$^{-2}$. The largest spatial scale detectable in the 610~MHz map is about 4\arcmin, which corresponds to a physical size of 1~Mpc. This indicates that a possible radio halo in \object{ZwCl~2341.1+0000} has a size $\gtrsim 1$~Mpc, or a very low surface brightness; or alternatively, it does not exist at all.

An optical R band image from the Wide Field Imaging (WFI) system at the MPG/ESO 2.2m telescope, showing the large-scale galaxy distribution around \object{ZwCl~2341.1+0000}, is shown in Fig.~\ref{fig:erben}. For more details about this image see \citeauthor{temple} (in prep.). The galaxy isodensity contours are also indicated in Figs.~\ref{fig:xray} and \ref{fig:erben}. The surface density of galaxies (per square arcmin) was computed using the photometric redshifts from the SDSS DR7 catalogs. Galaxies between $0.24+z_{\mathrm{err}} <z < 0.31-z_{\mathrm{err}}$ were selected from the catalogs, with $z_{\mathrm{err}}$ (typically between $0.02$ and $0.06$) the error in the photometric redshift. This assures that galaxies at approximately the distance of the cluster are selected while fore-/background galaxies are omitted. The galaxy distribution more or less follows  the X-ray emission. The fainter northeastern extension in the X-ray emission is associated with a galaxy filament extending roughly in the same direction. Several substructures are visible  within the main north-south galaxy structure.

\section{Spectral index \& equipartition magnetic field strength}
\label{sec:alphaandgauss}
\subsection{Spectral index}
\label{sec:spectra}

\begin{figure*}
    \begin{center}
      \includegraphics[angle = 90, trim=0cm 0cm 0cm 0cm,width=1.0\textwidth]{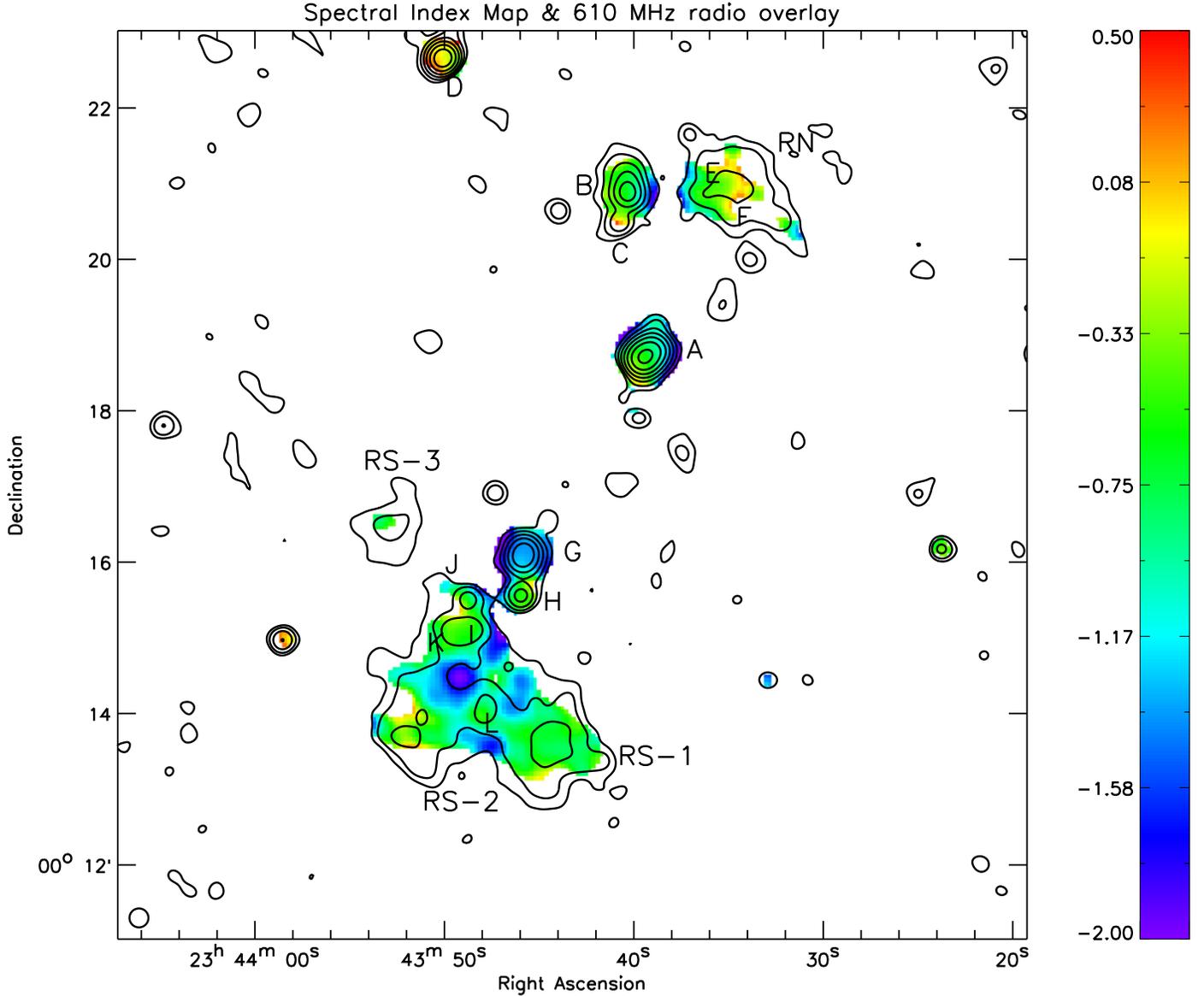}
       \end{center}
      \caption{Spectral index map between 610 and 241~MHz with a resolution of 18\arcsec. The black contours show the 610~MHz radio map convolved to a resolution of 15\arcsec. The radio contours are drawn at $[1, 2, 4, 8, 16, 32,  ...]  \times 0.224$ mJy beam$^{-1}$.}
            \label{fig:spix}
 \end{figure*}
\begin{figure}
    \begin{center}
      \includegraphics[angle = 90, trim=0cm 0cm 0cm 0cm,width=0.5\textwidth]{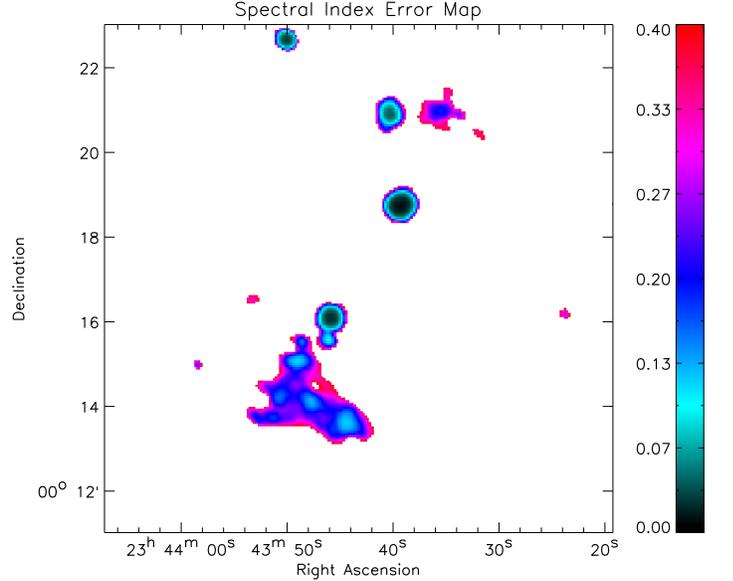}
       \end{center}
      \caption{Spectral index error map between 610 and 241~MHz.  The uncertainty of about $10\%$ in the absolute flux calibration, resulting in an additional error of $\pm 0.15$, is not included.}
             \label{fig:spix_error}
 \end{figure}

Radio spectra can be an important tool for understanding the origin of the relativistic electrons. We have created a spectral index map between 610 and 241~MHz, by convolving the 610~MHz image to the 241~MHz resolution, see Fig.~\ref{fig:spix}. We blanked pixels with a signal-to-noise ratio (SNR) below 3 in both maps.
The errors in the spectral index are displayed in Fig.~\ref{fig:spix_error} and are based on the rms noise in the images. 
Taking the uncertainty of about $10\%$ in the absolute flux calibration into account results in an additional systematic error in the spectral index of $\pm 0.15$. 

Spectral index variations are visible for RS and RN. At the location of compact sources (I, J, K, and L) embedded within the relic RS, the spectral index is relatively flat, around $-0.5$. A region with steep steep spectra is found in the northern region of RS-2. The spectral index of RN seems to steepen towards the east. However one has to be careful not to overinterpret these variations, as residual calibration errors may have resulted in artificial variations. Furthermore, differences in the UV-coverage can also have introduced spurious variations in the spectral index map.
For source A, there is an indication of spectral steepening towards the northwest consistent with the hypothesis that the core of the radio source is located southeast. The same is observed for source B, with steepening towards the north.

The integrated spectral index for RN (excluding the compact sources) is $-0.49 \pm 0.10 \pm 0.15$ , with the second error the previously reported systemic flux calibration error. The spectral index of $-0.49$ is consistent with the value of $-0.47 \pm 0.2$ measured by  \cite{2002NewA....7..249B}. For RS (not including RS-3), we find $-0.76 \pm 0.09 \pm 0.15$, somewhat steeper than the value of $-0.5 \pm 0.15$ reported by \citeauthor{2002NewA....7..249B} but consistent within the errors. The spectral indices were determined by fitting a simple power-law spectra through the integrated fluxes for RN and RS. The flux measurements in this work were combined with those measured from the images presented by \cite{2002NewA....7..249B}. The fluxes were determined in a fixed region, which was the same at all frequencies (without blanking). Compact sources were subtracted using the flux measurements at 610~MHz and a spectral index of $-0.5$.

\subsection{Equipartition magnetic field}
The presence of magnetic fields in the cluster on scales $\sim$1~Mpc is demonstrated from the observed synchrotron radiation from the two radio relics (RN and RS). The strength of the magnetic field can be estimated by assuming minimum energy densities in the radio sources. The minimum energy density (in units of erg cm$^{-3}$) is given by
\begin{equation}
u_{\mathrm{min}} =  \xi(\alpha, \nu_{1}, \nu_{2})(1+k)^{4/7}\nu_{0}^{-4\alpha/7}(1+z)^{(12-4\alpha)/7}I_{0}^{4/7}d^{-4/7} 
\label{eq:umin}
\end{equation}
with $\xi(\alpha, \nu_{1})$, a constant tabulated in \cite{2004IJMPD..13.1549G}, typically between $10^{-12}-10^{-14}$, $I_{0}$ the surface brightness (mJy arcsec$^{-2}$) at frequency $\nu_{0}$ (MHz), $d$ the depth of the source (kpc), and $k$ the ratio of the energy in relativistic protons to that in electrons (taken often as $k=1$ or $k=0$, behind a shock k is in the range $1-100$). A volume-filling factor of one has been assumed in the above equation. The equipartition magnetic field strength can then be calculated as
\begin{equation}
B_{\mathrm{eq}} = \left( \frac{24 \pi}{7} u_{\mathrm{min}} \right)^{1/2}\mbox{ .}
\label{eq:beq}
\end{equation}
This method calculates the synchrotron luminosity using a fixed high and low-frequency cutoff ($\nu_{1}$ and $\nu_{2}$). However, this is not entirely correct since the upper and lower limits should not be fixed during the integration because they dependent on the energy of the radiating electrons. Instead, low and high energy cutoffs for the particle distribution should be used \citep{1997A&A...325..898B, 2005AN....326..414B}. Taking this into account (and assuming that $\gamma_{\mathrm{min}} \ll \gamma_{\mathrm{max}}$, the energy boundaries indicated by the Lorentz factor), the revised equipartition magnetic field strength ($B^{\prime}_{\mathrm{eq}}$) is
\begin{equation}
B^{\prime}_{\mathrm{eq}} \sim 1.1 \gamma_{\mathrm{min}}^{(1+2\alpha)/(3-\alpha)} B_{\mathrm{eq}}^{7/(6-2\alpha)} \mbox{ .}
\label{eq:beqprime}
\end{equation}
Using a ratio of unity for the energy in relativistic protons to that in electrons (i.e., $k=1$) and applying the first of the previous two methods, we derive an equipartition magnetic field of $B_{\mathrm{eq}} = 0.59$ $\mu$Gauss for the northern relic ($\xi = 2.13 \times 10^{-12}$, $\nu_{1}=10$~MHz, $\nu_{2} = 10$~GHz), and $B_{\mathrm{eq}} = 0.55$ $\mu$Gauss for the southern relic ($\xi = 8.75 \times 10^{-13}$).
For the depth along the line of sight, we have taken the average of the major and minor axis of the relics. 
Using the lower and higher energy cutoff limits results in $B^{\prime}_{\mathrm{eq}}= 0.64$ $\mu$Gauss (RN) and  $0.93$ $\mu$Gauss (RS), assuming $\gamma_{\mathrm{min}} =100$. Using $\gamma_{\mathrm{min}} = 5000$ leads to $B^{\prime}_{\mathrm{eq}}= 0.66$ $\mu$Gauss and  $0.48$~$\mu$Gauss, respectively. The exact value for the lower cutoff is difficult to estimate. The Lorentz factor ($\gamma$) can be estimated according to $\gamma \sim 5 \times 10^{2}~\times\left( \nu\mbox{[MHz]}/ B_{\perp}\mbox{[$\mu$Gauss]} \right)$. The exact value for $\gamma_{\mathrm{min}}$ depends on the shape of the radio spectrum. Taking $k=100$ results in magnetic field strengths of about a factor of three higher than $k=1$.

\section{Discussion}

 \label{sec:discussion}
We have interpreted the diffuse radio emission in the periphery of \object{ZwCl~2341.1+0000} as a double radio relic, arising from outgoing shock fronts because of a cluster merger. This interpretation is based on (i) the location of the diffuse radio emission with respect to the X-ray emission, (ii) the presence of an elongated structure of galaxies in optical images, (iii) the orientation of the symmetry axis of the double relic perpendicular to the elongation axis of the X-ray and optical emission, (iv) the morphology of the X-ray emission, (v) the lack of a direct connection between the diffuse emission and the radio galaxies within the cluster, and (vi) the presence of head-tail galaxies, which are commonly found in merging clusters \citep[e.g.,][]{1994ApJ...423...94B}. This is all clear evidence that we are witnessing a merging system of subclusters where electrons are (re-)accelerated by large-scale shocks. Neither the Chandra nor the XMM-Newton X-ray images (\citeauthor{temple}) show any shock fronts at the location of the relics. However, both observations are probably too short to
see any sharp details in the low-brightness X-ray regions so
far from the cluster center.

\subsection{Alternative explanations}
 AGN also provides a source of relativistic electrons, and the question arises whether the diffuse radio emission could be related to an AGN. In particular, could RS and RN be the lobes of a giant radio galaxy located within the cluster? This seems unlikely, as (i) no host galaxy is visible located roughly halfway between RN and RS. The most likely candidate would then be source~A, but this source has the morphology of a compact head tail source, with no indication of a large-scale jet. The arm-length ratio of 1.9 between the two lobes and the core would also be quite extreme, because arm-length ratios are typically smaller than $1.5$ \citep[e.g.,][]{2008MNRAS.383..525K}. A faint 0.6 mJy source (RA~23$^{h}$~43$^{m}$~39.7$^{s}$,  DEC~00\degr~17\arcmin~55\farcs~0) is located below~A, resulting in a more reasonable arm-length ratio of about 1.3. But this source seems to be associated with a much more distant   $z_{\mathrm{phot}} =0.70$ galaxy. It could also be that the host galaxy is no longer active and therefore no central radio source is visible; however, in this case the spectral index of the lobes is expected to be steeper than about $-1$, which is not observed. (ii) This would make it one of the largest radio galaxies known with a size of 2.2~Mpc, but radio galaxies of this size are very rare. The largest known radio source is \object{J1420$-$0545}, with a 
projected size of 4.69~Mpc \citep{2008ApJ...679..149M}, and (iii) these giant radio galaxies do not reside in a high-density cluster environment. \cite{2002NewA....7..249B} came to the same conclusion that sources RN and RS are probably not related to a giant radio galaxy. 
Another possibility is that the electrons originate from some of the compact sources located in or around the regions with diffuse emission.
However, RN does not seem to be associated with the proposed head-tail source B, with the tail pointing north and RN located to the west. Source E and F are located within the diffuse emission but are unresolved. Source F would then be the most likely candidate because it is located at the distance of the cluster (source E is located behind the cluster with $z_{\mathrm{phot}} =0.45$). However, in that case we would expect some head-tail morphology, but the source is compact. 

RS-1 and RS-2 could be the radio lobes from source L, which coincides with an elliptical galaxy (probably a member of the cluster,  $z_{\mathrm{phot}} =0.30$). The radio source would then span a size of about 700~kpc. However the morphology of RS-1 and RS-2 is not entirely consistent with this scenario because no obvious lobe structure is visible. Furthermore there are no jets originating from L and the radio bridge connecting the two relics (RS-1 and RS-2) is very wide. In fact source L is completely embedded within the diffuse emission. The diffuse emission could also be associated with sources J, K, which could have supplied the radio plasma. However, no clear connection is visible although sources J and K are resolved. Both of these scenarios also fail to explain the presence of RS-3. An association with source G is also unlikely since it seems to be completely detached from the diffuse emission. 

\subsection{Comparison of spectral indices and magnetic field strengths with other double relics}

\cite{1997MNRAS.290..577R} report a spectral index of $\sim -1.1$ for the northern relic of A3667 (between 85.5, 408, and 843~MHz). For \object{RXC J1314.4$-$2515}, \cite{2007A&A...463..937V} report a spectral index of about $-1.4$ for both relics between 1400 and 610~MHz.  \cite{2009A&A...494..429B} report integrated spectral indices of $-1.5$ and $-1.3$ between 1400 and 325~MHz for A2345 and $-1.2$, $-1.3$ for A1240. Spectral steepening was observed away from the shock front towards the center of the cluster for one relic in A1240 (the data for the other relic was consistent with this trend) as well as for one of the relics in A2345. If the relics trace outward traveling shock fronts, spectral steepening towards the cluster center is expected because of spectral aging. Our low SNR in the 241~MHz map means that this cannot be tested for the double relics in \object{ZwCl~2341.1+0000}, although this prediction relies on several assumptions about the viewing angle, the magnetic field structure, and some assumptions about the merger geometry. A detailed merger scenario is presently unavailable from optical and X-ray observations. 

Our derived spectral indices of $-0.49$ (RN) and -0.76 (RS) are not particularly steep, although the errors in the spectral indices are relatively large. We also note that some large-scale variations in the background flux level are present, in particular east of the northern relic. By measuring the total flux within with an area just east of RN, we estimate that this could have potentially resulted in the flux loss of 10~mJy. If correct, this suggests a spectral indices of $\sim-0.6$ for RN. High-sensitivity observations, at for example 325~MHz, are needed to resolve this issue and create better spectral index maps.

Equipartition magnetic fields of about $1\mu$Gauss have been derived for several clusters having radio halos or relics \citep[e.g.,][]{2004IJMPD..13.1549G}. Our derived equipartition magnetic field strength of approximately $0.6$~$\mu$Gauss is comparable to values derived for other double relics. \cite{2006Sci...314..791B} derive a value of $0.5-3.0$~$\mu$Gauss for the double relics in \object{A3376}, and \cite{2009A&A...494..429B} derived field strengths of $1.0-2.5$~$\mu$Gauss for \object{A1240} and $0.8-2.9$~$\mu$Gauss for \object{A2345}. 

Equipartition field strength should be used with caution, not only for their dependence on merely guessed properties of the electron spectrum, but also for the assumption of equipartition between relativistic particle and magnetic
field energies, which may or may not be established by physical processes in
the synchrotron emitting region. Faraday rotation and inverse Compton (IC) scattering of CMB photons by relativistic electrons in the relics can also give an independent estimate of the magnetic field strength. Both techniques give similar results (within a factor of 10) to the derived equipartition magnetic field strength, indicating that the assumptions made are reasonable \citep[see][and references therein]{2004IJMPD..13.1549G,2008SSRv..134...93F, 2009A&A...494..429B}.

\subsection{Origin of the double relic}


The presence of Mpc scale radio emission in the periphery of the
cluster requires an acceleration mechanism for emitting 
relativistic particles.
This is naturally provided by the diffusive shock acceleration
mechanism \citep[DSA;][]{1977DoSSR.234R1306K, 1977ICRC...11..132A, 1978MNRAS.182..147B, 1978MNRAS.182..443B, 1978ApJ...221L..29B, 1983RPPh...46..973D, 1987PhR...154....1B, 1991SSRv...58..259J, 2001RPPh...64..429M} via the Fermi-I 
process~\citep{1998A&A...332..395E, 2001ApJ...562..233M}. In this scenario
the synchrotron spectral index, $\alpha$,  of the relativistic 
electrons is determined by the slope, $q$, of the underlying 
CR distribution function,
which in turn depends on the shock Mach number, ${\mathcal M}$ \citep{1959flme.book.....L, 2002ASSL..272....1S}. The
relevant expressions (in linear theory) are
\begin{equation}
\alpha =-\frac{q-3}{2} \mbox{, } q=\frac{4}{1-{\mathcal M}^{-2}}  \mbox{ or, }
\alpha =-\frac{3{\mathcal M}^{-2}+1}{2 - 2{\mathcal M}^{-2}} \mbox{ .}
\end{equation}
For strong shocks, $\mathcal{M} \gg 1$, resulting in a flat spectral index
of about $-0.5$. 

However, the high-energy electrons responsible for the synchrotron emission
have a finite lifetime given by
\begin{equation}
t_{\mathrm{age}} = 1060 \frac{B^{0.5}}{B^2 + B_{\mathrm{IC}}^2}  [{(1 + z)\nu_{b}}]^{-0.5}   \mbox{ [Myr]} \mbox{ ,}
\end{equation}
with $B$ the magnetic field strength and $B_{\mathrm{IC}}= 3.25(1+z)^2$ the
equivalent magnetic field strength of the microwave background both in
units of $\mu$Gauss \citep[e.g.,][]{1980ARA&A..18..165M, 2001AJ....122.1172S}. 
We have no information about the break frequency $\nu_{b}$ (expressed
in GHz) but assume a reasonable value of $\gtrsim 1$~GHz, because the
spectral indices between 241 and 610~MHz are not particularly steep.
Then for a magnetic field strength of 0.6~$\mu$Gauss $t_{\mathrm{age}}
\lesssim 50$~Myr.

In the post-shock flow, the accelerated particles are most likely
trapped by the strong turbulent field generated by the shock itself.
Then, the extent of the region within which diffuse radio emission is
expected to be visible can be roughly estimated by taking the product
of the particles cooling time and the shock speed~\citep{2001ApJ...562..233M}.
This typically gives a few hundred kpc, roughly consistent with the
thickness of the sources observed in Fig.~\ref{fig:xray}.  The lateral extent, on the
other hand, is related to the size and strength of the shock.
In addition, the energy losses mean that the longer the distances from the acceleration site 
(shock front), the lower the energy cutoff appearing in the particle distribution
function. As it turns out, the combination of the particles distribution
functions at different locations has a slope that is steeper by one
unit with respect to the one at the acceleration site \citep{2002NewA....7..249B}. As a result,
although the spectrum is flat at the acceleration site ($\alpha \sim-0.5$), 
we expect to measure a spectral index closer to minus one. Higher values than this (i.e., flatter radio spectra) can either indicate a continuous acceleration in the post-shock
region, incomplete shock acceleration model, a systematic error in the
calculation of the spectral index, or a combination thereof.

Our spectral spectral indices are relatively flat and marginally consistent
with the above description in which the radio emitting electrons are
shock-accelerated. Such flat spectral indices have also been found for other
merger related relics. An example is the relic in \object{Abell 2256}.
\cite{2008A&A...489...69B} measures a spectral index of $\sim -0.5$ for
some parts of the relic between 338 and 365~MHz. The integrated
spectral index for the relic was found to be $-0.72 \pm 0.02$, similar
to that in \object{ ZwCl~2341.1+0000}.

High mach number shocks are required for efficient particle
acceleration~\citep{1987PhR...154....1B}.  Numerical simulations indicate the
development of different types of shocks during structure formation,
including external accretion shocks, as well as merger and flow
shocks, both of which are internal to a galaxy
cluster~\citep{2000ApJ...542..608M}.  These shocks differ in their respective
Mach number, \citep[e.g.,][]{2000ApJ...542..608M, 2002MNRAS.337..199M, 2003ApJ...593..599R,
 2006MNRAS.367..113P,2008ApJ...689.1063S}. External accretion shocks,
which process the low-density, unshocked IGM, have $\mathcal{M} \gg 1$,
which result in flat spectral indices at the acceleration site of
about $-0.5$. Although some level of diffuse radio emission is
expected there, the surface brightness is too low to be detected by
current facilities, owing to the expected low density of both magnetic
field energy and CR particles there \citep{2001ApJ...562..233M, 2008MNRAS.391.1511H,
 2008MNRAS.385.1211P}.  Nevertheless, these shocks could possibly
shine and be detected in gamma rays
~\citep{2000Natur.405..156L,2002MNRAS.337..199M,2003MNRAS.342.1009M,2003ApJ...585..128K,2007ApJ...667L...1M}.  Internal shocks produced by accretion through filaments are weaker than the external accretion
shocks but still strong enough to produce flat spectra at the shock front.
In addition, unlike external accretion shocks, they occur in a high
enough density environment to be detected by
current radio facilities.  Finally, binary merger shocks are weak and inefficient in
the initial stages of the impact. As the shocks propagate outward into
the lower density environment, however, they steepen and evolve into
high Mach number shocks. Therefore, these shocks too 
are viable candidates for the origin of the observed radio emission. Also, a weak shock with $\mathcal{M} \sim 2.3$ seems to be able to produce a peripheral radio relic in A521 \citep{2008A&A...486..347G}, so it may be that the Mach numbers do not need to be that high for efficient particle acceleration.

While, in principle, according to simulations
double relics are allowed in the filament-shock-accretion picture
discussed above, there is no reason why they should be 
symmetric with respect to any particular axis. Instead, a symmetric configuration with respect to the X-ray elongation axis
arises quite naturally in the binary merger
picture~\citep{1999ApJ...518..603R}.
Thus the morphological and symmetry properties of the source as
they emerge from the combined X-ray and radio image seem to strongly 
suggest that the double radio relic is associated with a binary merger 
scenario.


\section{ Conclusions}

\label{sec:conclusion}

We have presented low-frequency radio observations of the merging cluster \object{ ZwCl~2341.1+0000} at 610, 241, and 157~MHz taken with the GMRT. The radio maps show two diffuse structures to the north and south of the cluster, which we classify as a double radio relic, where the particles are accelerated by the DSA mechanism in an outward moving shock. Our interpretation is different from \cite{2002NewA....7..249B}, which proposed that the radio emission originated from cosmic shocks in the IGM. The relics are located along the elongated X-ray axis (i.e, the merger axis). Their orientation is perpendicular to this axis and the radio emission straddles the outer boundary of the X-ray emission. Several possible head-tail sources are also found within the cluster. 

The derived galaxy distribution shows an elongated structure consisting out of several substructures, and a galaxy filament extending towards the northeast that seems to be connected to the main structure. This extension is also visible in the Chandra X-ray images. The radio spectral indices found are relatively flat, $-0.49 \pm 0.18$ for the northern relic and $-0.76 \pm 0.17$ for the southern relic. The derived equipartition magnetic field strength is $\sim0.6$~$\mu$Gauss, comparable to values derived for other double relics. The two radio relics on both sides of the cluster are probably outward traveling shocks caused by a major merger event. High SNR radio spectral index maps, together with more detailed optical and X-ray analyses of the cluster, will be needed to further test the merger scenario, in particular whether radial spectral steepening of the relics is observed towards the cluster center. This would be evidence of an outward traveling shock.

\begin{acknowledgements}
We would like to thank the anonymous referee for the useful comments. We would like to thank T.~E. Clarke for her help with the GMRT proposal. We thank the staff of the GMRT who have made these observations possible. GMRT is run by the National Centre for Radio Astrophysics of the Tata Institute of Fundamental Research. RJvW acknowledges funding from the Royal Netherlands Academy of 
Arts and Sciences.

Funding for the SDSS and SDSS-II has been provided by the Alfred P. Sloan Foundation, the Participating Institutions, the National Science Foundation, the U.S. Department of Energy, the National Aeronautics and Space Administration, the Japanese Monbukagakusho, the Max Planck Society, and the Higher Education Funding Council for England. The SDSS Web Site is http://www.sdss.org/. The SDSS is managed by the Astrophysical Research Consortium for the Participating Institutions. The Participating Institutions are the American Museum of Natural History, Astrophysical Institute Potsdam, University of Basel, University of Cambridge, Case Western Reserve University, University of Chicago, Drexel University, Fermilab, the Institute for Advanced Study, the Japan Participation Group, Johns Hopkins University, the Joint Institute for Nuclear Astrophysics, the Kavli Institute for Particle Astrophysics and Cosmology, the Korean Scientist Group, the Chinese Academy of Sciences (LAMOST), Los Alamos National Laboratory, the Max-Planck-Institute for Astronomy (MPIA), the Max-Planck-Institute for Astrophysics (MPA), New Mexico State University, Ohio State University, University of Pittsburgh, University of Portsmouth, Princeton University, the United States Naval Observatory, and the University of Washington.

    This research has made use of the NASA/IPAC Extragalactic Database (NED), which is operated by the Jet Propulsion Laboratory, California Institute of  Technology, under contract with the National Aeronautics and Space Administration. This research has made use of the VizieR catalogue access tool, CDS, Strasbourg, France. 

\end{acknowledgements}

\bibliographystyle{aa}
\bibliography{12287b}
 \Online
\clearpage
 \begin{appendix} 
 \section{Compact sources at 610~MHz and optical counterparts}
\label{sec:compact}
\begin{figure}
    \begin{center}
      \includegraphics[angle = 90, trim =0cm 0cm 0cm 0cm,width=0.5\textwidth]{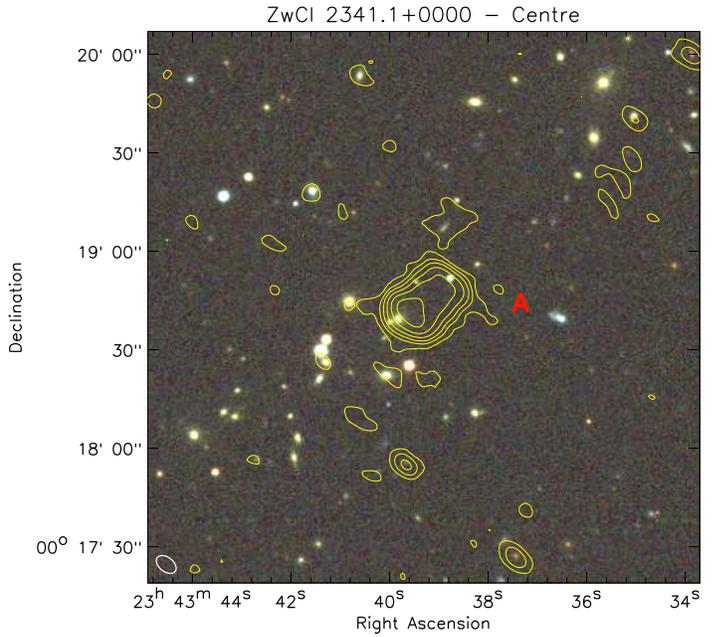}
      \end{center}
      \caption{SDSS DR7 image color image (\emph{g}, \emph{r}, and \emph{i} bands) of the cluster center overlaid with radio contours. The 610~MHz contour levels are drawn at  $\sqrt{[1, 8, 32, 128, ...]}  \times 84$ $\mu$Jy beam$^{-1}$. The beam size of $6.9\arcsec \times 4.3\arcsec$ is indicated in the bottom left corner.}
\label{centreoptical}
 \end{figure}

\begin{figure}
    \begin{center}
      \includegraphics[angle = 90, trim =0cm 0cm 0cm 0cm,width=0.5\textwidth]{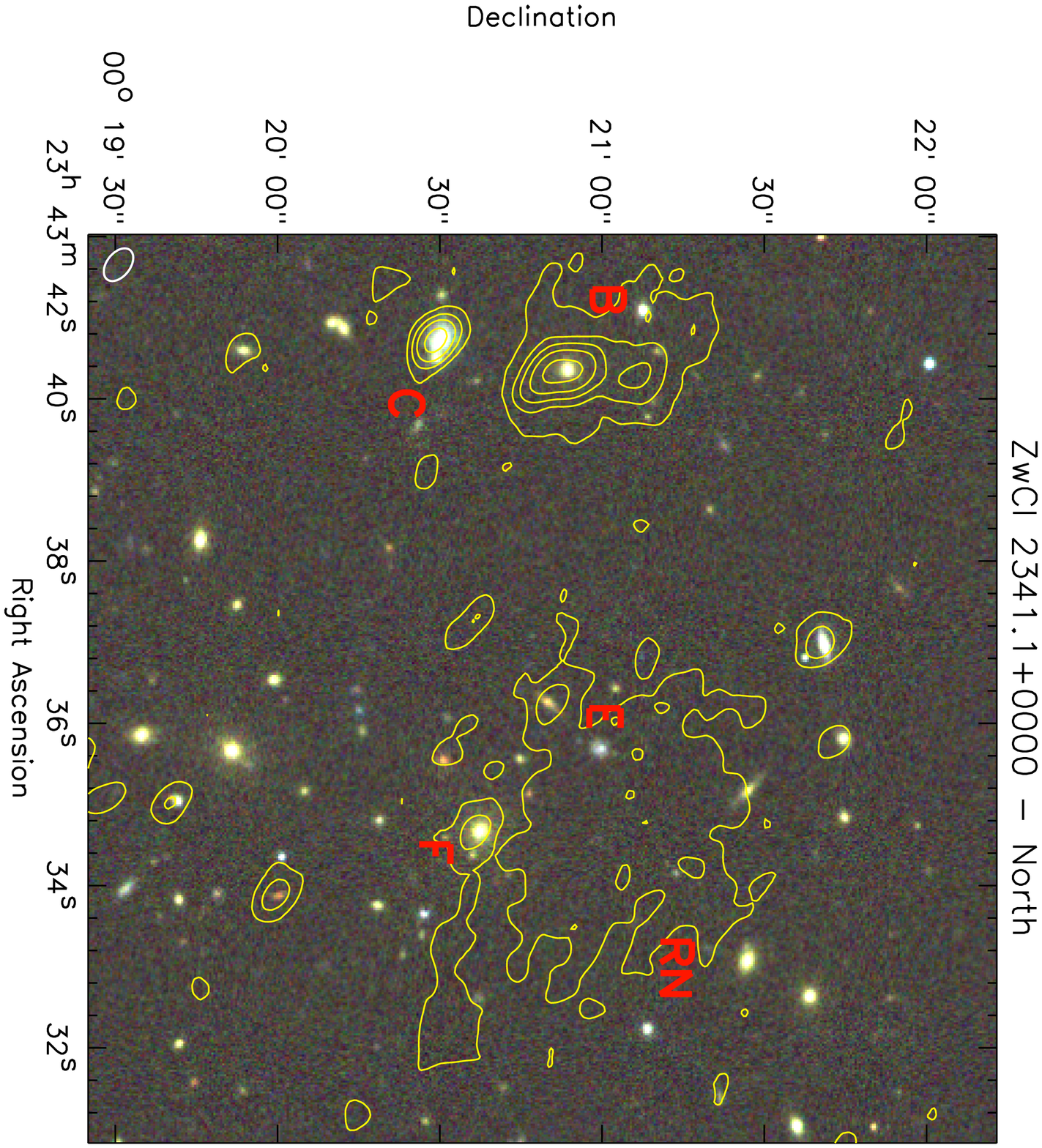}
       \end{center}
      \caption{SDSS DR7  image color image (\emph{g}, \emph{r}, and \emph{i} bands) of the northern part of the cluster overlaid with 610~MHz radio contours. Contours are drawn at the same levels as in Fig.~\ref{centreoptical}.}
             \label{fig:north_optical}
 \end{figure}

\begin{figure*}
    \begin{center}
      \includegraphics[angle = 90, trim=0cm 0cm 0cm 0cm,width=1.\textwidth]{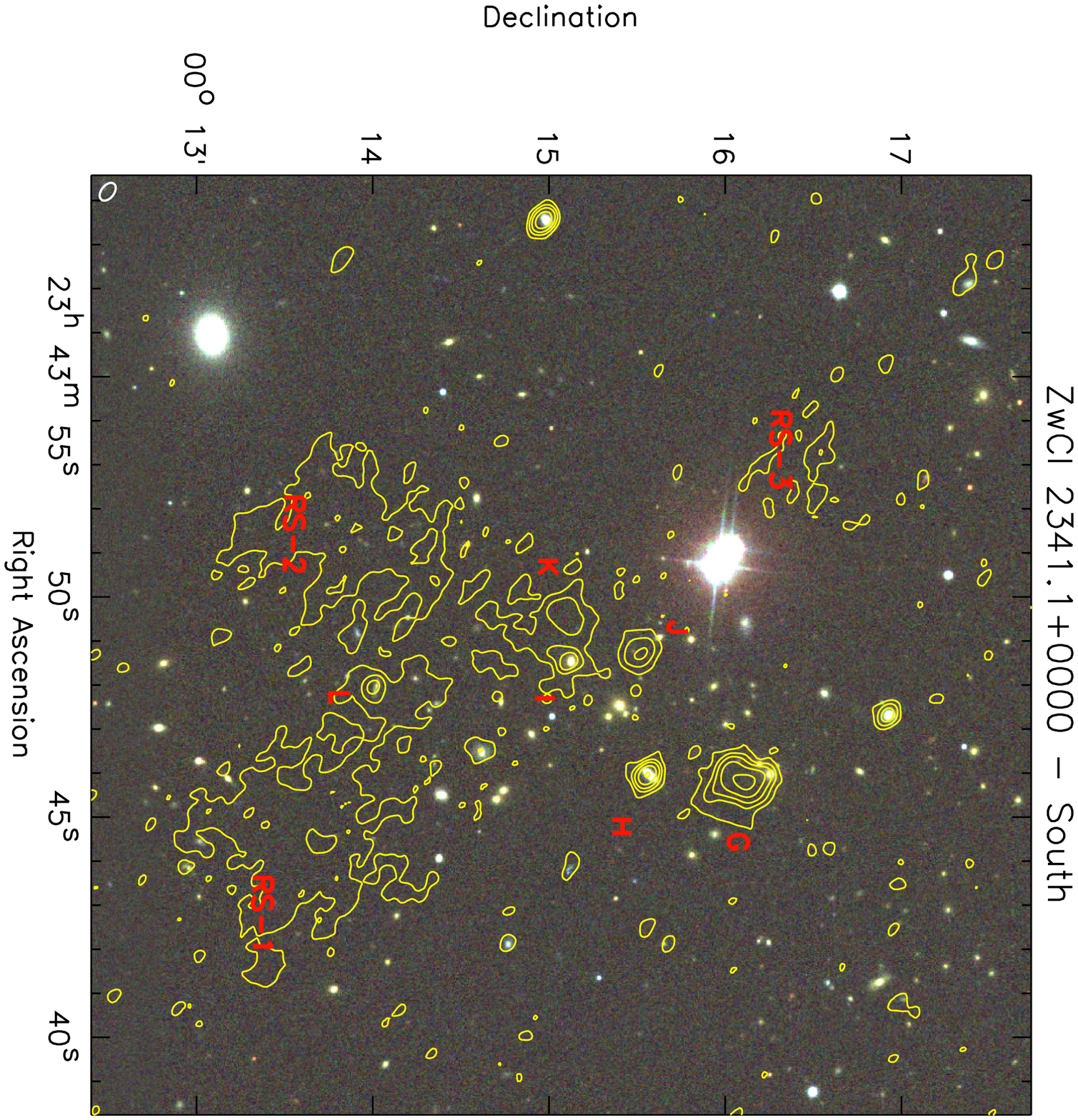}
       \end{center}
      \caption{SDSS DR7  image color image (\emph{g}, \emph{r}, and \emph{i} bands) of the southern part of the cluster overlaid with 610~MHz radio contours. Contours are drawn at the same levels as in Fig.~\ref{centreoptical}.}
             \label{fig:south_optical}
 \end{figure*}

To identify optical counterparts for the radio sources we have overlaid the 610~MHz radio contours on SDSS images, see Figs. \ref{centreoptical},  \ref{fig:north_optical}, and \ref{fig:south_optical}.  

Source A is resolved, 13\arcsec~by 19\arcsec, with the brightest part of the emission located to the southeast. The resolution is insufficient to classify the source, but the morphology is consistent with a head-tail source. A spectral index gradient is also observed towards the northwest (Sect. \ref{sec:spectra}). An optical counterpart is visible to the southwest of the center of the radio source, close to the peak of the radio emission (see Fig.~\ref{centreoptical}). The photometric redshift ($z_{\mathrm{phot}}$) of this galaxy is $0.32$, consistent with being a cluster member. The E/S0 galaxy (mag$_{\mathrm{r}}$ = 18.65) has a close companion (mag$_{\mathrm{r}}$ = 21.12) about 3\arcsec~to the southwest. The spectral index of A is $-0.92$, between 157 and 1400~MHz, typical for a radio galaxy. The source therefore does not seem to be directly related to the diffuse emission within the cluster, as suggested by \cite{2002NewA....7..249B}.

Source B has a peculiar morphology, resembling a head-tail galaxy. A mag$_{\mathrm{r}}$ = 18.65 E/S0 counterpart is located at $z_{\mathrm{phot}} =0.29$. The direction of the tail suggest the galaxy is falling in from the north towards the cluster center. The spectral index of the source is $-0.76$ (between 241 and 610~MHz). Other sources north of the cluster center are sources C, E, and F. Source C has a blue star-forming galaxy (mag$_{\mathrm{r}} = 17.87$) as an optical counterpart at $z=0.261$ (spectroscopic redshift). The SDSS DR6 spectrum of this galaxy shows strong emission lines (in particular H$\alpha$). A close companion is located about 4\arcsec~to the west. The E/S0 mag$_{\mathrm{r}}$ = 18.06  counterpart of source F has a spectroscopic redshift of 0.269, and the SDSS spectrum shows a strong Balmer break. The mag$_{\mathrm{r}}$= 20.11 optical counterpart of E is located at $z_{\mathrm{phot}} =0.45$ and may therefore not be associated with the cluster.

To the south of the cluster center we have sources G, H, I, J, K, and L. Source G is probably a head-tail source, with the tail pointing south, indicating the galaxy is also falling towards the cluster center. The spectral index $\alpha_{241}^{610}$ of $-1.53$ is consistent with such an identification, and there is a hint of spectral steepening towards the south. The counterpart of source G is an E/S0 mag$_{\mathrm{r}}$ = 18.35 galaxy (located at $z_{\mathrm{phot}} =0.35$) with two close companions. The counterpart of source H is a blue mag$_{\mathrm{r}}$ = 18.02 galaxy at  $z_{\mathrm{phot}} =0.28$. The spectral index $\alpha_{241}^{610}$ of $\sim$$-1.08$, quite steep for a star-forming galaxy \citep[e.g.,][]{1993ApJ...405..498W,2006ApJ...645..186T, 2007A&A...463..519B}, this indicates that an AGN may also be present. The counterpart of source I is an E/S0 mag$_{\mathrm{r}}$ = 18.45 galaxy located at  $z_{\mathrm{phot}} =0.26$. The counterpart of source L is an E/S0 mag$_{\mathrm{r}}$ = 19.84 galaxy with $z_{\mathrm{phot}} =0.30$ and $\alpha_{241}^{610} \sim -0.5$. The galaxy is located roughly in the middle between RS-1 and RS-2. Radio sources J and K are diffuse with sizes of about 10\arcsec. Source K has no counterpart, while source J has a possible mag$_{\mathrm{r}}$= 22.23,  $z_{\mathrm{phot}} =0.18$ counterpart. However, J and K could also be the two lobes of a distant radio galaxy.

 \end{appendix}
\end{document}